\documentclass[twoside,english,onecolumn]{IEEEtran}
\usepackage[T1]{fontenc}
\usepackage[latin9]{inputenc}
\usepackage{babel}
\usepackage{verbatim}
\usepackage{float}
\usepackage{amsthm}
\usepackage{amsmath}
\usepackage{amssymb}
\usepackage{setspace}
\usepackage{esint}
\PassOptionsToPackage{normalem}{ulem}
\usepackage{ulem}
\usepackage[unicode=true,
 bookmarks=false,
 breaklinks=false,pdfborder={0 0 0},backref=false,colorlinks=false]
 {hyperref}
\hypersetup{pdftitle={Approximately Optimal Trajectory Tracking for an Uncertain Nonlinear System},
 pdfauthor={Rushikesh Kamalapurkar},
 pdfkeywords={Adaptive Dynamic Programming, Approximate Dynamic Programming, Nonlinear Control, Optimal Control},
 pdfstartview= FitBH}
\usepackage{breakurl}

\makeatletter

\floatstyle{ruled}
\newfloat{algorithm}{tbp}{loa}
\providecommand{\algorithmname}{Algorithm}
\floatname{algorithm}{\protect\algorithmname}

\theoremstyle{definition}
\newtheorem{assumption}{Assumption}
  \theoremstyle{plain}
  \newtheorem{lem}{\protect\lemmaname}
  \theoremstyle{plain}
  \newtheorem{thm}{\protect\theoremname}

\usepackage{cite}
\makeatother

\providecommand{\lemmaname}{Lemma}
\providecommand{\theoremname}{Theorem}

\begin{document}

\title{Approximately Optimal Trajectory Tracking for Continuous Time Nonlinear
Systems%
\thanks{Rushikesh Kamalapurkar and Warren Dixon are with the Department of
Mechanical and Aerospace Engineering, University of Florida, Gainesville,
FL, USA. Email: \{rkamalapurkar, wdixon\}@ufl{}.edu.%
}%
\thanks{Shubhendu Bhasin is with the Department of Electrical Engineering,
Indian Institute of Technology, Delhi, India. email: sbhasin@ee.iitd.ac.in. %
}%
\thanks{Huyen Dinh is with the Department of Mechanical Engineering, University
of Transport and Communications, Hanoi, Vietnam. email: huyendtt214@gmail.com. %
}%
\thanks{This research is supported in part by NSF award numbers 0901491, 1161260,
1217908, ONR grant number N00014-13-1-0151, and a contract with the
AFRL Mathematical Modeling and Optimization Institute. Any opinions,
findings and conclusions or recommendations expressed in this material
are those of the authors and do not necessarily reflect the views
of the sponsoring agency.%
}}

\author{Rushikesh Kamalapurkar, Huyen Dinh, Shubhendu Bhasin, and Warren
Dixon}
\maketitle
\begin{abstract}
Approximate dynamic programming has been investigated and used as
a method to approximately solve optimal regulation problems. However,
the extension of this technique to optimal tracking problems for continuous
time nonlinear systems has remained a non-trivial open problem. The
control development in this paper guarantees ultimately bounded tracking
of a desired trajectory, while also ensuring that the controller converges
to an approximate optimal policy.
\end{abstract}

\section{Introduction}

Reinforcement learning (RL) is a concept that can be used to enable
an agent to learn optimal policies from interaction with the environment.
The objective of the agent is to learn the policy that maximizes or
minimizes a cumulative long term reward. Almost all RL algorithms
use some form of generalized policy iteration (GPI). GPI is a set
of two simultaneous interacting processes, policy evaluation and policy
improvement. Starting with an estimate of the state value function
and an admissible policy, policy evaluation makes the estimate consistent
with the policy and policy improvement makes the policy greedy with
respect to the value function. These algorithms exploit the fact that
the optimal value function satisfies Bellman's principle of optimality
\cite{Kirk2004,Sutton1998}. 

The principle of optimality leads to a wide range of algorithms that
focus on finding solutions to the Bellman equation (BE) or approximations
of the BE. For discrete time systems, BE-based policy evaluation methods
do not require a model of the environment, and hence, have been central
to the development of RL \cite{Sutton1998}. Approximate dynamic programming
(ADP) consists of algorithms that facilitate the solution of the approximate
BE for problems with a continuous state space or an infinite discrete
state space by utilizing a function approximation structure to approximate
the state value function \cite{Bertsekas2007}.%

When applied to continuous time systems the principle of optimality
leads to the Hamilton-Jacobi-Bellman (HJB) equation which is the continuous
time counterpart of the BE \cite{Doya2000}. Similar to discrete time
ADP, continuous time ADP approaches aim at finding approximate solutions
to the HJB equation. Various methods to solve this problem are proposed
in \cite{Beard1997,Abu-Khalaf2002,Vrabie2009,Vamvoudakis2010,Bhasin.Kamalapurkar.ea2013a,Jiang.Jiang2012,Zhang.Luo.ea2009}
and the references therein. An infinite horizon regulation problem
with a quadratic cost function is the most common problem considered
in ADP literature. For these problems, function approximation techniques
can be used to approximate the value function because it is time-invariant. 

Approximation techniques like NNs are commonly used in ADP literature
for value function approximation. ADP-based approaches are presented
in results such as \cite{Dierks2010,Zhang.Cui.ea2011} to address
the tracking problem for continuous time systems, where the value
function, and the controller presented are time-varying functions
of the tracking error. However, for the infinite horizon optimal control
problem, time does not lie on a compact set, and NNs can only approximate
functions on a compact domain. Thus, it is unclear how a NN with the
tracking error as an input can approximate the time-varying value
function and controller. 

For discrete time systems, several approaches have been developed
to address the tracking problem. Park et.al. \cite{Park1996} use
generalized back-propagation through time to solve a finite horizon
tracking problem that involves offline training of NNs. An ADP-based
approach is presented in \cite{Dierks2009a} to solve an infinite
horizon optimal tracking problem where the desired trajectory is assumed
to depend on the system states. Greedy heuristic dynamic programming
based algorithms are presented in results such as \cite{Zhang2008a,Luo.Liang2011,Wang.Liu.ea2012}
which transform the nonautonomous system into an autonomous system,
and approximate convergence of the sequence of value functions to
the optimal value function is established. However, these results
lack an accompanying stability analysis.

In this result, the tracking error and the desired trajectory both
serve as inputs to the NN. This makes the controller in (\ref{eq:control})
fundamentally different from previous results, in the sense that a
different HJB equation must be solved and its solution, i.e. the feedback
component of the controller, is a time-varying function of the tracking
error. In particular, this paper addresses the technical obstacles
that result from the time-varying nature of the optimal control problem
by including the partial derivative of the value function with respect
to the desired trajectory in the HJB equation, and by using a system
transformation to convert the problem into a time-invariant optimal
control problem in such a way that the resulting value function is
a time-invariant function of the transformed states, and hence, lends
itself to approximation using a NN. A Lyapunov-based analysis is used
to prove ultimately bounded tracking and that the controller converges
to the approximate optimal policy.

\section{Formulation of time-invariant optimal control problem\label{sec:Formulation-of-stationary}}

Consider a class of nonlinear control affine systems 
\[
\dot{x}=f\left(x\right)+g\left(x\right)u,
\]
where $x\in\mathbb{R}^{n}$ is the state, and $u\in\mathbb{R}^{m}$
is the control input. The functions $f:\mathbb{R}^{n}\rightarrow\mathbb{R}^{n}$
and $g:\mathbb{R}^{n}\rightarrow\mathbb{R}^{n\times m}$ are locally
Lipschitz, where $f\left(0\right)=0$, and the solution of the system
is unique for any bounded initial condition $x\left(t_{0}\right)\in\mathbb{R}^{n}$
and control $u$. The control objective is to track a bounded continuously
differentiable signal $x_{d}:\mathbb{R}_{\geq0}\rightarrow\mathbb{R}^{n}$.
To quantify this objective, a tracking error is defined as $e\triangleq x-x_{d}$.
The open-loop tracking error dynamics can then be written as
\begin{equation}
\dot{e}=f\left(x\right)+g\left(x\right)u-\dot{x}_{d}.\label{eq:errordyn}
\end{equation}

The following assumptions are made to facilitate the formulation of
an approximate optimal tracking controller.
\begin{assumption}
\label{Apsinv} The function $g$ is bounded and has full column rank,
and the function $g^{+}:\mathbb{R}^{n}\rightarrow\mathbb{R}^{m\times n}$
defined as $g^{+}\triangleq\left(g^{T}g\right)^{-1}g^{T}$ is bounded
and locally Lipschitz. 
\end{assumption}

\begin{assumption}
\label{Ah}The desired trajectory is bounded such that $\left\Vert x_{d}\right\Vert \leq d\in\mathbb{R}$,
and there exists a locally Lipschitz function $h_{d}:\mathbb{R}^{n}\rightarrow\mathbb{R}^{n}$
such that $\dot{x}_{d}=h_{d}\left(x_{d}\right)$ and $h_{d}\left(0\right)=0$.
\end{assumption}
The steady-state control policy $u_{d}:\mathbb{R}^{n}\to\mathbb{R}^{m}$
corresponding to the desired trajectory $x_{d}$ is 
\begin{equation}
u_{d}\left(x_{d}\right)=g_{d}^{+}\left(h_{d}\left(x_{d}\right)-f_{d}\right),\label{eq:ud}
\end{equation}
where $g_{d}^{+}\triangleq g^{+}\left(x_{d}\right)$ and $f_{d}\triangleq f\left(x_{d}\right)$,
respectively. To transform the time-varying optimal control problem
into a time-invariant optimal control problem, a new concatenated
state $\zeta\in\mathbb{R}^{2n}$ is defined as \cite{Zhang2008a}
\begin{equation}
\zeta\triangleq\left[e^{T},x_{d}^{T}\right]^{T}.\label{eq:zeta}
\end{equation}
Based on (\ref{eq:errordyn}) and Assumption \ref{Ah}, the time derivative
of (\ref{eq:zeta}) can be expressed as 
\begin{equation}
\dot{\zeta}=F\left(\zeta\right)+G\left(\zeta\right)\mu,\label{eq:AffineDyn}
\end{equation}
where the functions $F:\mathbb{R}^{2n}\to\mathbb{R}^{2n}$, $G:\mathbb{R}^{2n}\to\mathbb{R}^{2n\times m}$,
and the control $\mu\in\mathbb{R}^{m}$ are defined as
\begin{gather}
F\left(\zeta\right)\triangleq\left[\begin{gathered}f\left(e+x_{d}\right)-h_{d}\left(x_{d}\right)+g\left(e+x_{d}\right)u_{d}\left(x_{d}\right)\\
h_{d}\left(x_{d}\right)
\end{gathered}
\right],\quad G\left(\zeta\right)\triangleq\begin{bmatrix}g\left(e+x_{d}\right)\\
0
\end{bmatrix},\quad\mu\triangleq u-u_{d}.\label{eq:mu}
\end{gather}
Local Lipschitz continuity of $f$ and $g$, the fact that $f\left(0\right)=0$,
and Assumption \ref{Ah} imply that $F\left(0\right)=0$ and $F$
is locally Lipschitz. The objective of the optimal control problem
is to design a policy $\mu^{*}:\mathbb{R}^{2n}\to\mathbb{R}^{m}\in\Psi$
that minimizes the cost functional $V:\mathbb{R}_{\geq0}\to\mathbb{R}_{\geq0}$
defined as 
\begin{equation}
V\left(t\right)\triangleq\int_{t}^{\infty}r\left(\zeta\left(\rho\right),\mu^{*}\left(\rho\right)\right)d\rho,\label{eq:V}
\end{equation}
subject to the dynamic constraints in (\ref{eq:AffineDyn}), where
$\Psi$ is the set of admissible policies \cite{Beard1997}, and $r:\mathbb{R}^{2n}\times\mathbb{R}^{m}\to\mathbb{R}_{\geq0}$
is the local cost defined as 
\begin{equation}
r\left(\zeta,\mu\right)\triangleq\zeta^{T}\overline{Q}\zeta+\mu{}^{T}R\mu.\label{eq:localcost}
\end{equation}
In (\ref{eq:localcost}), $R\in\mathbb{R}^{m\times m}$ is a positive
definite symmetric matrix of constants, and $\overline{Q}\in\mathbb{R}^{2n\times2n}$
is defined as 
\begin{equation}
\overline{Q}\triangleq\begin{bmatrix}Q & 0_{n\times n}\\
0_{n\times n} & 0_{n\times n}
\end{bmatrix},\label{eq:qbar}
\end{equation}
where $Q\in\mathbb{R}^{n\times n}$ is a positive definite symmetric
matrix of constants with the minimum eigenvalue $\underline{q}\in\mathbb{R}_{>0}$,
and $0_{n\times n}\in\mathbb{R}^{n\times n}$ is a matrix of zeros.
For brevity of notation, let $\left(\cdot\right)^{\prime}$ denote
$\partial\left(\cdot\right)/\partial\zeta$

\section{Approximate optimal solution}

Assuming that the minimizing policy exists, the HJB equation for the
optimal control problem can be written as 
\begin{equation}
H^{*}=V^{*\prime}\left(F+G\mu^{*}\right)+r\left(\zeta,\mu^{*}\right)=0,\label{eq:HJB*}
\end{equation}
where $H^{*}$ denotes the Hamiltonian, $V^{*}:\mathbb{R}^{2n}\to\mathbb{R}_{\geq0}$
denotes the optimal value function, and $\mu^{*}:\mathbb{R}^{2n}\to\mathbb{R}^{m}$
denotes the optimal policy. For the local cost in (\ref{eq:localcost})
and the dynamics in (\ref{eq:AffineDyn}), the optimal policy can
be obtained in closed-form as \cite{Kirk2004} 
\begin{equation}
\mu^{*}=-\frac{1}{2}R^{-1}G^{T}\left(V^{*\prime}\right)^{T},\label{eq:mu*}
\end{equation}
assuming that the optimal value function satisfies $V^{*}\in C^{1}$
and $V^{*}\left(0\right)=0.$ 

On any compact set $\chi\subset\mathbb{R}^{2n}$, the value function
$V^{*}$ can be represented using a NN with $N$ neurons as
\begin{equation}
V^{*}\left(\zeta\right)=W^{T}\sigma\left(\zeta\right)+\epsilon\left(\zeta\right),\label{eq:V*NN}
\end{equation}
where $W\in\mathbb{R}^{N}$ is the ideal weight matrix bounded above
by a known positive constant $\bar{W}\in\mathbb{R}$ in the sense
that $\left\Vert W\right\Vert \leq\bar{W}$, $\sigma:\mathbb{R}^{2n}\rightarrow\mathbb{R}^{N}$
is a bounded continuously differentiable nonlinear activation function,
and $\epsilon:\mathbb{R}^{2n}\rightarrow\mathbb{R}$ is the function
reconstruction error such that $\sup_{\zeta\in\mathbb{\chi}}\left|\epsilon\left(\zeta\right)\right|\leq\bar{\epsilon}$
and\textbf{ }$\sup_{\zeta\in\mathbb{\chi}}\left|\epsilon^{\prime}\left(\zeta\right)\right|\leq\bar{\epsilon^{\prime}}$,
where $\bar{\epsilon}\in\mathbb{R}$ and $\bar{\epsilon^{\prime}}\in\mathbb{R}$
are positive constants \cite{Hornik1990,Lewis2002}.

Using $\left(\ref{eq:mu*}\right)$ and $\left(\ref{eq:V*NN}\right)$
the optimal policy can be represented as
\begin{equation}
\mu^{*}=-\frac{1}{2}R^{-1}G^{T}\left(\sigma^{\prime T}W+\epsilon^{\prime T}\right).\label{eq:mu*NN}
\end{equation}
Based on $\left(\ref{eq:V*NN}\right)$ and $\left(\ref{eq:mu*NN}\right)$,
the NN approximations to the optimal value function and the optimal
policy are given by
\begin{gather}
\hat{V}=\hat{W}_{c}^{T}\sigma,\qquad\mu=-\frac{1}{2}R^{-1}G^{T}\sigma^{\prime T}\hat{W}_{a},\label{eq:muhat}
\end{gather}
where $\hat{W}_{c}\in\mathbb{R}^{N}$ and $\hat{W}_{a}\in\mathbb{R}^{N}$
are estimates of the ideal neural network weights $W$. The use of
two separate sets of weight estimates $\hat{W}_{a}$ and $\hat{W}_{c}$
for $W$ is motivated by the fact that the Bellman error is linear
with respect to the value function weight estimates and nonlinear
with respect to the policy weight estimates. Use of a separate set
of weight estimates for the value function facilitates least squares-based
adaptive updates.

The controller is obtained from (\ref{eq:ud}), (\ref{eq:mu}), and
(\ref{eq:muhat}) as 
\begin{equation}
u=-\frac{1}{2}R^{-1}G^{T}\sigma^{\prime T}\hat{W}_{a}+g_{d}^{+}\left(h_{d}-f_{d}\right).\label{eq:control}
\end{equation}

Using the approximations $\mu$ and $\hat{V}$ for $\mu^{*}$ and
$V^{*}$ in (\ref{eq:HJB*}), respectively, the approximate Hamiltonian
$\hat{H}$ can be obtained as $\hat{H}=\hat{V}^{\prime}\left(F+G\mu\right)+r\left(\zeta,\mu\right)$.
Using (\ref{eq:HJB*}), the error between the approximate and the
optimal Hamiltonian, called the Bellman Error $\delta\in\mathbb{R}$,
is given in a measurable form by
\begin{align}
\delta & \triangleq\hat{H}-H^{*}=\hat{V}^{\prime}\left(F+G\mu\right)+r\left(\zeta,\mu\right).\label{eq:delta}
\end{align}
The value function weights are updated to minimize $\intop_{0}^{t}\delta^{2}\left(\rho\right)d\rho$
using a normalized least squares update law with an exponential forgetting
factor as \cite{Ioannou1996}
\begin{align}
\dot{\hat{W}}_{c} & =-\eta_{c}\Gamma\frac{\omega}{1+\nu\omega^{T}\Gamma\omega}\delta,\label{eq:WcHatdot}\\
\dot{\Gamma} & =-\eta_{c}\left(-\lambda\Gamma+\Gamma\frac{\omega\omega^{T}}{1+\nu\omega^{T}\Gamma\omega}\Gamma\right),\label{eq:Gammadot}
\end{align}
where $\nu,\eta_{c}\in\mathbb{R}$ are positive adaptation gains,
$\omega\in\mathbb{R}^{N}$ is defined as $\omega\triangleq\sigma^{\prime}\left(F+G\mu\right)$,
and $\lambda\in\left(0,1\right)$ is the forgetting factor for the
estimation gain matrix $\Gamma\in\mathbb{R}^{N\times N}$. The policy
weights are updated to follow the critic weights as
\begin{align}
\overset{\cdot}{\hat{W}}_{a} & =-\eta_{a1}\left(\hat{W}_{a}-\hat{W}_{c}\right)-\eta_{a2}\hat{W}_{a},\label{eq:WaHatdot}
\end{align}
where $\eta_{a1},\eta_{a2}\in\mathbb{R}$ are positive adaptation
gains.

Using (\ref{eq:HJB*}), (\ref{eq:delta}), and (\ref{eq:WcHatdot}),
an unmeasurable form of the BE can be written as
\begin{equation}
\delta=-\tilde{W}_{c}^{T}\omega+\frac{1}{4}\tilde{W}_{a}^{T}\mathcal{G}_{\sigma}\tilde{W}_{a}+\frac{1}{4}\epsilon^{\prime}\mathcal{G}\epsilon^{\prime T}+\frac{1}{2}W^{T}\sigma^{\prime}\mathcal{G}\epsilon^{\prime T}-\epsilon^{\prime}F,\label{eq:deltaUnm}
\end{equation}
where $\mathcal{G}\triangleq GR^{-1}G^{T}$ and $\mathcal{G}_{\sigma}\triangleq\sigma^{\prime}GR^{-1}G^{T}\sigma^{\prime T}$.
The weight estimation errors for the value function and the policy
are defined as $\tilde{W}_{c}\triangleq W-\hat{W}_{c}$ and $\tilde{W}_{a}\triangleq W-\hat{W}_{a}$,
respectively. Using (\ref{eq:deltaUnm}), the weight estimation error
dynamics for the value function are
\begin{equation}
\dot{\tilde{W}}_{c}=-\eta_{c}\Gamma\psi\psi^{T}\tilde{W}_{c}+\eta_{c}\Gamma\frac{\omega}{1+\nu\omega^{T}\Gamma\omega}\Bigl(\frac{1}{4}\tilde{W}_{a}^{T}\mathcal{G}_{\sigma}\tilde{W}_{a}+\frac{1}{4}\epsilon^{\prime}\mathcal{G}\epsilon^{\prime T}+\frac{1}{2}W^{T}\sigma^{\prime}\mathcal{G}\epsilon^{\prime T}-\epsilon^{\prime}F\Bigr),\label{eq:WcDyn}
\end{equation}
where $\psi\triangleq\frac{\omega}{\sqrt{1+\nu\omega^{T}\Gamma\omega}}\in\mathbb{R}^{N}$
is the regressor vector. 
\begin{assumption}
\label{APE}The regressor $\psi:\mathbb{R}_{\geq0}\to\mathbb{R}^{N}$
is persistently exciting (PE). Thus, there exist $T,\underline{\psi}>0$
such that $\underline{\psi}I\leq\intop_{t}^{t+T}\psi\left(\tau\right)\psi\left(\tau\right)^{T}d\tau.$ 
\end{assumption}
The dynamics in (\ref{eq:WcDyn}) can be regarded as a perturbed form
of the nominal system $\dot{\tilde{W}}_{c}=-\eta_{c}\Gamma\psi\psi^{T}\tilde{W}_{c}.$
Using Assumption \ref{APE} and \cite[Corollary 4.3.2]{Ioannou1996}
it can be concluded that 
\begin{equation}
\underline{\varphi}I_{N\times N}\leq\Gamma\left(t\right)\leq\overline{\varphi}I_{N\times N},\:\forall t\in\mathbb{R}_{\geq0}\label{eq:GammaBound}
\end{equation}
where $\overline{\varphi},\underline{\varphi}\in\mathbb{R}$ are constants
such that $0<\underline{\varphi}<\overline{\varphi}$. Based on (\ref{eq:GammaBound}),
the regressor vector can be bounded as
\begin{equation}
\left\Vert \psi\left(t\right)\right\Vert \leq\frac{1}{\sqrt{\nu\underline{\varphi}}},\:\forall t\in\mathbb{R}_{\geq0}.\label{eq:psibound}
\end{equation}
Using Assumptions \ref{Apsinv} and \ref{Ah} and the fact that on
any compact set, there exists a positive constant $L_{F}\in\mathbb{R}$
such that $\left\Vert F\right\Vert \leq L_{F}\left\Vert \zeta\right\Vert ,$
the following bounds are developed to aid the subsequent stability
analysis: 
\begin{gather}
\left\Vert \left(\frac{\epsilon^{\prime}}{4}+\frac{W^{T}\sigma}{2}'\right)\mathcal{G}\epsilon^{\prime T}\right\Vert +\overline{\epsilon^{\prime}}L_{F}\left\Vert x_{d}\right\Vert \leq\iota_{1},\quad\left\Vert \mathcal{G}_{\sigma}\right\Vert \leq\iota_{2},\quad\left\Vert \epsilon^{\prime}\mathcal{G}\epsilon^{\prime T}\right\Vert \leq\iota_{3},\quad\left\Vert \frac{1}{2}W^{T}\mathcal{G}_{\sigma}+\frac{1}{2}\epsilon^{\prime}\mathcal{G}\sigma^{\prime T}\right\Vert \leq\iota_{4},\nonumber \\
\left\Vert \frac{1}{4}\epsilon^{\prime}\mathcal{G}\epsilon^{\prime T}+\frac{1}{2}W^{T}\sigma^{\prime}\mathcal{G}\epsilon^{\prime T}\right\Vert \leq\iota_{5},\label{eq:bounds}
\end{gather}
where $\iota_{1},\iota_{2},\iota_{3},\iota_{4},\iota_{5}\in\mathbb{R}$
are positive constants.

\section{Stability Analysis}

The contribution in the previous section was the development of a
transformation that enables the optimal policy and the optimal value
function to be expressed as a time-invariant function of $\zeta$.
The use of this transformation presents a challenge in the sense that
the optimal value function, which is used as the Lyapunov function
for the stability analysis, is not a positive definite function of
$\zeta$ because the matrix $\overline{Q}$ is positive semidefinite.
In this section, this technical obstacle is addressed by exploiting
the fact that the time-invariant optimal value function $V^{*}:\mathbb{R}^{2n}\rightarrow\mathbb{R}$
can be interpreted as a time-varying map $V_{t}^{*}:\mathbb{R}^{n}\times\mathbb{R}_{\geq0}\rightarrow\mathbb{R}$,
such that 
\begin{equation}
V_{t}^{*}\left(e,t\right)=V^{*}\left(\left[\begin{array}{c}
e\\
x_{d}\left(t\right)
\end{array}\right]\right)\label{eq:Vt}
\end{equation}
for all $e\in\mathbb{R}^{2n}$ and for all $t\in\mathbb{R}_{\geq0}$.
Specifically, the time-invariant form facilitates the development
of the approximate optimal policy, whereas the equivalent time-varying
form can be shown to be a positive definite and decresent function
of the tracking error. In the following, Lemma \ref{lem:decresent}
and Lemma \ref{lem:V*pd} are used to prove that $V_{t}^{*}:\mathbb{R}^{n}\times\mathbb{R}_{\geq0}\rightarrow\mathbb{R}$
is positive definite and decresent, and hence, a candidate Lyapunov
function. Theorem \ref{thm:main_thm} then states the main result
of the paper.
\begin{lem}
\label{lem:decresent}Let $D\subseteq\mathbb{R}^{n}$ contain the
origin and let $\Xi:D\times\mathbb{R}_{\geq0}\rightarrow\mathbb{R}_{\geq0}$
be positive definite. If $\Xi\left(x,t\right)$ is bounded, uniformly
in $t$, for all bounded $x$ and if $x\longmapsto\Xi\left(x,t\right)$
is continuous, uniformly in $t$, then $\Xi$ is decresent in $D$.\end{lem}
\begin{IEEEproof}
Since $\Xi\left(x,t\right)$ is bounded, uniformly in $t$, $\sup_{t\in\mathbb{R}_{\geq0}}\left\{ \Xi\left(x,t\right)\right\} $
exists and is unique for all bounded $x$. Let the function $\alpha:D\rightarrow\mathbb{R}_{\geq0}$
be defined as 
\begin{equation}
\alpha\left(x\right)\triangleq\sup_{t\in\mathbb{R}_{\geq0}}\left\{ \Xi\left(x,t\right)\right\} .\label{eq:sup}
\end{equation}
Since $x\to\Xi\left(x,t\right)$ is continuous, uniformly in $t$,
$\forall\varepsilon>0$, $\exists\varsigma\left(x\right)>0$ such
that $\forall y\in D$, 
\begin{equation}
d_{D\times\mathbb{R}_{\geq0}}\left(\left(x,t\right),\left(y,t\right)\right)<\varsigma\left(x\right)\implies d_{\mathbb{R}\geq0}\left(\Xi\left(x,t\right),\Xi\left(y,t\right)\right)<\varepsilon,\label{eq:epdel1}
\end{equation}
where $d_{M}\left(\cdot,\cdot\right)$ denotes the standard Euclidean
metric on the metric space $M$. By the definition of $d_{M}\left(\cdot,\cdot\right)$,
$d_{D\times\mathbb{R}_{\geq0}}\left(\left(x,t\right),\left(y,t\right)\right)=d_{D}\left(x,y\right).$
Using (\ref{eq:epdel1}), 
\begin{equation}
d_{D}\left(x,y\right)<\varsigma\left(x\right)\implies\left|\Xi\left(x,t\right)-\Xi\left(y,t\right)\right|<\varepsilon.\label{eq:epdel2}
\end{equation}
Given the fact that $\Xi$ is positive, (\ref{eq:epdel2}) implies
$\Xi\left(x,t\right)<\Xi\left(y,t\right)+\varepsilon$ and $\Xi\left(y,t\right)<\Xi\left(x,t\right)+\varepsilon$
which from (\ref{eq:sup}) implies $\alpha\left(x\right)<\alpha\left(y\right)+\varepsilon$
and $\alpha\left(y\right)<\alpha\left(x\right)+\varepsilon$, and
hence, from (\ref{eq:epdel2}), $d_{D}\left(x,y\right)<\varsigma\left(x\right)\implies\left|\alpha\left(x\right)-\alpha\left(y\right)\right|<\varepsilon.$
Since $\Xi$ is positive definite, (\ref{eq:sup}) can be used to
conclude $\alpha\left(0\right)=0.$ Thus, $\Xi$ is bounded above
by a continuous positive definite function, and hence, is decresent
in $D$.\end{IEEEproof}
\begin{lem}
\label{lem:V*pd}Let $B_{a}$ denote a closed ball around the origin
with the radius $a\in\mathbb{R}_{>0}$. The optimal value function
$V_{t}^{*}:\mathbb{R}^{n}\times\mathbb{R}_{\geq0}\rightarrow\mathbb{R}$
satisfies the following properties\begin{subequations}
\begin{align}
\ensuremath{V_{t}^{*}\left(e,t\right)} & \geq\underline{v}\left(\left\Vert e\right\Vert \right),\label{eq:V*pd1}\\
\ensuremath{V_{t}^{*}\left(0,t\right)} & =0,\label{eq:V*pd2}\\
V_{t}^{*}\left(e,t\right) & \leq\overline{v}\left(\left\Vert e\right\Vert \right),\label{eq:V*decrescent}
\end{align}
 \end{subequations}$\forall t\in\mathbb{R}_{\geq0}$ and $\forall e\in B_{a}$
where $\underline{v}:\left[0,a\right]\rightarrow\mathbb{R}_{\geq0}$
and $\overline{v}:\left[0,a\right]\rightarrow\mathbb{R}_{\geq0}$
are class $\mathcal{K}$ functions.\end{lem}
\begin{IEEEproof}
Based on the definitions in (\ref{eq:V})-(\ref{eq:qbar}) and (\ref{eq:Vt}),
\begin{gather}
V_{t}^{*}\left(e,t\right)=\intop_{t}^{\infty}\left(e^{T}\left(\rho\right)Qe\left(\rho\right)+\mu^{*T}\left(\rho\right)R\mu^{*}\left(\rho\right)\right)d\rho\geq V_{e}\left(e\right),\:\forall t\in\mathbb{R}_{\geq0},\label{eq:V*}
\end{gather}
where $V_{e}\left(e\right)\triangleq\intop_{t}^{\infty}\left(e^{T}\left(\rho\right)Qe\left(\rho\right)\right)d\rho$
is a positive definite function. Lemma 4.3 in \cite{Khalil2002} can
be invoked to conclude that there exists a class $\mathcal{K}$ function
$\underline{v}:\left[0,a\right]\rightarrow\mathbb{R}_{\geq0}$ such
that $\underline{v}\left(\left\Vert e\right\Vert \right)\leq V_{e}\left(e\right)$,
which along with (\ref{eq:V*}), implies (\ref{eq:V*pd1}).

From (\ref{eq:V*}), $V^{*}\left(\left[0,x_{d}^{T}\right]^{T}\right)=\intop_{t}^{\infty}\left(\mu^{*T}\left(\rho\right)R\mu^{*}\left(\rho\right)\right)d\rho,$
with the minimizer $\mu^{*}\left(t\right)=0,\forall t\in\mathbb{R}_{\geq0}.$
Furthermore, $V^{*}\left(\left[0,x_{d}^{T}\right]^{T}\right)$ is
the cost incurred when starting with $e=0$ and following the optimal
policy thereafter for any arbitrary desired trajectory $x_{d}$ (cf.
Section 3.7 of \cite{Sutton1998}). Substituting $x\left(t_{0}\right)=x_{d}\left(t_{0}\right)$,
$\mu\left(t_{0}\right)=0$ and (\ref{eq:ud}) in (\ref{eq:AffineDyn})
indicates that $\dot{e}\left(t_{0}\right)=0.$ Thus, when starting
from $e=0$, the zero policy satisfies the dynamic constraints in
(\ref{eq:AffineDyn}). Furthermore, the optimal cost is $V^{*}\left(\left[0,x_{d}^{T}\right]^{T}\right)=0,\:\forall\left\Vert x_{d}\right\Vert <d$
which, from (\ref{eq:Vt}), implies (\ref{eq:V*pd2}).

Admissibility of the optimal policy implies that $V^{*}\left(\zeta\right)$
is bounded for all bounded $\zeta$. Since the desired trajectory
is bounded, $V_{t}^{*}\left(e,t\right)$ is bounded, uniformly in
$t$, for all bounded $e$. To establish that $e\longmapsto V_{t}^{*}\left(e,t\right)$
is continuous, uniformly in $t$, let $\chi_{e_{o}}\subset\mathbb{R}^{n}$
be a compact set containing $e_{o}$. Since $x_{d}$ is bounded, $x_{d}\in\chi_{x_{d}}$,
where $\chi_{x_{d}}\subset\mathbb{R}^{n}$ is compact. Since $V^{*}:\mathbb{R}^{2n}\to\mathbb{R}_{\geq0}$
is continuous, and $\chi_{e_{o}}\times\chi_{x_{d}}\subset\mathbb{R}^{2n}$
is compact, $V^{*}$ is uniformly continuous on $\chi_{e_{o}}\times\chi_{x_{d}}$.
Thus, $\forall\varepsilon>0,$ $\exists\varsigma>0$, such that $\forall\left[e_{o}^{T},x_{d}^{T}\right]^{T},\left[e_{1}^{T},x_{d}^{T}\right]^{T}\in\chi_{e_{o}}\times\chi_{x_{d}}$,
$d_{\chi_{e_{o}}\times\chi_{x_{d}}}\left(\left[e_{o}^{T},x_{d}^{T}\right]^{T},\left[e_{1}^{T},x_{d}^{T}\right]^{T}\right)<\varsigma\implies d_{\mathbb{R}}\left(V^{*}\left(\left[e_{o}^{T},x_{d}^{T}\right]^{T}\right),V^{*}\left(\left[e_{1}^{T},x_{d}^{T}\right]^{T}\right)\right)<\varepsilon.$
Thus, for each $e_{o}\in\mathbb{R}^{n}$, there exists a $\varsigma>0$
independent of $x_{d}$, that establishes the continuity of $e\longmapsto V^{*}\left(\left[e^{T},x_{d}^{T}\right]^{T}\right)$
at $e_{o}$. Thus, $e\longmapsto V^{*}\left(\left[e^{T},x_{d}^{T}\right]^{T}\right)$
is continuous, uniformly in $x_{d}$, and hence, using (\ref{eq:Vt}),
$e\longmapsto V_{t}^{*}\left(e,t\right)$ is continuous, uniformly
in $t$. Using Lemma \ref{lem:decresent} and (\ref{eq:V*pd1}) and
(\ref{eq:V*pd2}), there exists a positive definite function $\alpha:\mathbb{R}^{n}\rightarrow\mathbb{R}_{\geq0}$
such that $V_{t}^{*}\left(e,t\right)<\alpha\left(e\right),\:\forall\left(e,t\right)\in\mathbb{R}^{n}\times\mathbb{R}_{\geq0}$.
Lemma 4.3 in \cite{Khalil2002} indicates that there exists a class
$\mathcal{K}$ function $\overline{v}:\left[0,a\right]\rightarrow\mathbb{R}_{\geq0}$
such that $\alpha\left(e\right)\leq\overline{v}\left(\left\Vert e\right\Vert \right)$,
which implies (\ref{eq:V*decrescent}). \end{IEEEproof}
\begin{lem}
\label{lem:WaBound} Let $Z\triangleq\begin{bmatrix}e^{T} & \tilde{W}_{c}^{T} & \tilde{W}_{a}^{T}\end{bmatrix}^{T},$
and let $\chi\in\mathbb{R}^{n+2N}$ be a compact set such that $Z\left(\tau\right)\in\chi$,
for all $\tau\in[t,t+T]$. Then, the actor weights and the tracking
errors satisfy 
\begin{gather}
-\inf_{\tau\in\left[t,t+T\right]}\left\Vert e\left(\tau\right)\right\Vert ^{2}\leq-\varpi_{0}\sup_{\tau\in\left[t,t+T\right]}\left\Vert e\left(\tau\right)\right\Vert ^{2}+\varpi_{1}T^{2}\sup_{\tau\in\left[t,t+T\right]}\left\Vert \tilde{W}_{a}\left(\tau\right)\right\Vert ^{2}+\varpi_{2}\label{eq:ebound}\\
-\inf_{\tau\in\left[t,t+T\right]}\left\Vert \tilde{W}_{a}\left(\tau\right)\right\Vert ^{2}\leq-\varpi_{3}\sup_{\tau\in\left[t,t+T\right]}\left\Vert \tilde{W}_{a}\left(\tau\right)\right\Vert ^{2}+\varpi_{4}\inf_{\tau\in\left[t,t+T\right]}\left\Vert \tilde{W}_{c}\left(\tau\right)\right\Vert ^{2}+\varpi_{5}\sup_{\tau\in\left[t,t+T\right]}\left\Vert e\left(\tau\right)\right\Vert ^{2}+\varpi_{6},\label{eq:Wabound}
\end{gather}
where

\textup{$\varpi_{0}=\frac{\left(1-6nT^{2}L_{F}^{2}\right)}{2},$ $\varpi_{1}=\frac{3n}{4}\sup_{t}\left\Vert gR^{-1}G^{T}\sigma^{\prime T}\right\Vert ^{2}$,
$\varpi_{2}=\frac{3n^{2}T^{2}\left(dL_{F}+\sup_{t}\left\Vert gg_{d}^{+}\left(h_{d}-f_{d}\right)-\frac{1}{2}gR^{-1}G^{T}\sigma^{\prime T}W-h_{d}\right\Vert \right)^{2}}{n},$ }

\textup{$\varpi_{3}=\frac{\left(1-6N\left(\eta_{a1}+\eta_{a2}\right)^{2}T^{2}\right)}{2},$
$\varpi_{4}=\frac{6N\eta_{a1}^{2}T^{2}}{\left(1-6N\left(\eta_{c}\overline{\varphi}T\right)^{2}/\left(\nu\underline{\varphi}\right)^{2}\right)},$
$\varpi_{5}=\frac{18\left(\eta_{a1}N\eta_{c}\overline{\varphi}\bar{\epsilon^{\prime}}L_{F}T^{2}\right)^{2}}{\nu\underline{\varphi}\left(1-6N\left(\eta_{c}\overline{\varphi}T\right)^{2}/\left(\nu\underline{\varphi}\right)^{2}\right)},$ }

\textup{$\varpi_{6}=\frac{18\left(N\eta_{a1}\eta_{c}\overline{\varphi}\left(\bar{\epsilon^{\prime}}L_{F}d+\iota_{5}\right)T^{2}\right)^{2}}{\nu\underline{\varphi}\left(1-6N\left(\eta_{c}\overline{\varphi}T\right)^{2}/\left(\nu\underline{\varphi}\right)^{2}\right)}+3N\left(\eta_{a2}\overline{W}T\right)^{2}.$}\end{lem}
\begin{IEEEproof}
Using the definition of the controller in (\ref{eq:muhat}), the tracking
error dynamics can be expressed as
\[
\dot{e}=f+\frac{1}{2}gR^{-1}G^{T}\sigma'^{T}\tilde{W}_{a}+gg_{d}^{+}\left(h_{d}-f_{d}\right)-\frac{1}{2}gR^{-1}G^{T}\sigma'^{T}W-h_{d}.
\]
On any compact set, the tracking error derivative can be bounded above
as
\[
\left\Vert \dot{e}\right\Vert \leq L_{F}\left\Vert e\right\Vert +L_{W}\left\Vert \tilde{W}_{a}\right\Vert +L_{e},
\]
where $L_{e}=L_{F}\left\Vert x_{d}\right\Vert +\left\Vert gg_{d}^{+}\left(h_{d}-f_{d}\right)-\frac{1}{2}gR^{-1}G^{T}\sigma'^{T}W-h_{d}\right\Vert $
and $L_{W}=\frac{1}{2}\left\Vert gR^{-1}G^{T}\sigma'^{T}\right\Vert $.
Using the fact that $e$ and $\tilde{W}_{a}$ are continuous functions
of time, on the interval $\left[t,t+T\right]$, the time derivative
of $e$ can be bounded as 
\[
\left\Vert \dot{e}\right\Vert \leq L_{F}\sup_{\tau\in\left[t,t+T\right]}\left\Vert e\left(\tau\right)\right\Vert +L_{W}\sup_{\tau\in\left[t,t+T\right]}\left\Vert \tilde{W}_{a}\left(\tau\right)\right\Vert +L_{e}.
\]
Since the infinity norm is less than the 2-norm, the derivative of
the $j^{th}$ component of $\dot{e}$ is bounded as 
\[
\dot{e}_{j}\leq L_{F}\sup_{\tau\in\left[t,t+T\right]}\left\Vert e\left(\tau\right)\right\Vert +L_{W}\sup_{\tau\in\left[t,t+T\right]}\left\Vert \tilde{W}_{a}\left(\tau\right)\right\Vert +L_{e}.
\]
Thus, the maximum and the minimum value of $e_{j}$ are related as
\[
\sup_{\tau\in\left[t,t+T\right]}\left|e_{j}\left(\tau\right)\right|\leq\inf_{\tau\in\left[t,t+T\right]}\left|e_{j}\left(\tau\right)\right|+\left(L_{F}\sup_{\tau\in\left[t,t+T\right]}\left\Vert e\left(\tau\right)\right\Vert +L_{W}\sup_{\tau\in\left[t,t+T\right]}\left\Vert \tilde{W}_{a}\left(\tau\right)\right\Vert +L_{e}\right)T.
\]
Squaring the above expression and using the inequality $\left(x+y\right)^{2}\leq2x^{2}+2y^{2}$
\[
\sup_{\tau\in\left[t,t+T\right]}\left|e_{j}\left(\tau\right)\right|^{2}\leq2\inf_{\tau\in\left[t,t+T\right]}\left|e_{j}\left(\tau\right)\right|^{2}+2\left(L_{F}\sup_{\tau\in\left[t,t+T\right]}\left\Vert e\left(\tau\right)\right\Vert +L_{W}\sup_{\tau\in\left[t,t+T\right]}\left\Vert \tilde{W}_{a}\left(\tau\right)\right\Vert +L_{e}\right)^{2}T^{2}.
\]
Summing over $j$, and using the the facts that $\sup_{\tau\in\left[t,t+T\right]}\left\Vert e\left(\tau\right)\right\Vert ^{2}\leq\sum_{j=1}^{n}\sup_{\tau\in\left[t,t+T\right]}\left|e_{j}\left(\tau\right)\right|^{2}$
and $\inf_{\tau\in\left[t,t+T\right]}\sum_{j=1}^{n}\left|e_{j}\left(\tau\right)\right|^{2}\leq\inf_{\tau\in\left[t,t+T\right]}\left\Vert e\left(\tau\right)\right\Vert ^{2}$,
\[
\sup_{\tau\in\left[t,t+T\right]}\left\Vert e\left(\tau\right)\right\Vert ^{2}\leq2\inf_{\tau\in\left[t,t+T\right]}\left\Vert e\left(\tau\right)\right\Vert ^{2}+2\left(L_{F}\sup_{\tau\in\left[t,t+T\right]}\left\Vert e\left(\tau\right)\right\Vert ^{2}+L_{W}\sup_{\tau\in\left[t,t+T\right]}\left\Vert \tilde{W}_{a}\left(\tau\right)\right\Vert ^{2}+L_{e}\right)^{2}nT^{2}.
\]
Using the inequality $\left(x+y+z\right)^{2}\leq3x^{2}+3y^{2}+3z^{2}$,
(\ref{eq:ebound}) is obtained.

Using a similar procedure on the dynamics for $\tilde{W}_{a}$, 
\begin{multline}
-\inf_{\tau\in\left[t,t+T\right]}\left\Vert \tilde{W}_{a}\left(\tau\right)\right\Vert ^{2}\leq-\frac{\left(1-6N\left(\eta_{a1}+\eta_{a2}\right)^{2}T^{2}\right)}{2}\sup_{\tau\in\left[t,t+T\right]}\left\Vert \tilde{W}_{a}\left(\tau\right)\right\Vert ^{2}\\
+3N\eta_{a1}^{2}\sup_{\tau\in\left[t,t+T\right]}\left\Vert \tilde{W}_{c}\left(\tau\right)\right\Vert ^{2}T^{2}+3N\eta_{a2}^{2}W^{2}T^{2}.\label{eq:1}
\end{multline}
Similarly, the dynamics for $\tilde{W}_{c}$ yield 
\begin{multline}
\sup_{\tau\in\left[t,t+T\right]}\left\Vert \tilde{W}_{c}\left(\tau\right)\right\Vert ^{2}\leq\frac{2}{\left(1-\frac{6N\eta_{c}^{2}\overline{\varphi}^{2}T^{2}}{\nu^{2}\underline{\varphi}^{2}}\right)}\inf_{\tau\in\left[t,t+T\right]}\left\Vert \tilde{W}_{c}\left(\tau\right)\right\Vert ^{2}\\
+\frac{6NT^{2}\eta_{c}^{2}\overline{\varphi}^{2}\bar{\epsilon'}^{2}L_{F}^{2}}{\nu\underline{\varphi}\left(1-\frac{6N\eta_{c}^{2}\overline{\varphi}^{2}T^{2}}{\nu^{2}\underline{\varphi}^{2}}\right)}\sup_{\tau\in\left[t,t+T\right]}\left\Vert e\left(\tau\right)\right\Vert ^{2}+\frac{6NT^{2}\eta_{c}^{2}\overline{\varphi}^{2}\left(\bar{\epsilon'}L_{F}d+\iota_{5}\right)^{2}}{\nu\underline{\varphi}\left(1-\frac{6N\eta_{c}^{2}\overline{\varphi}^{2}T^{2}}{\nu^{2}\underline{\varphi}^{2}}\right)}.\label{eq:2}
\end{multline}
Substituting (\ref{eq:2}) into (\ref{eq:1}), the (\ref{eq:Wabound})
can be obtained.%
\end{IEEEproof}
\begin{lem}
\label{lem:WcBound}Let $Z\triangleq\begin{bmatrix}e^{T} & \tilde{W}_{c}^{T} & \tilde{W}_{a}^{T}\end{bmatrix}^{T},$
and let $\chi\in\mathbb{R}^{n+2N}$ be a compact set such that $Z\left(\tau\right)\in\chi$,
for all $\tau\in[t,t+T]$. Then, the critic weights satisfy
\[
-\intop_{t}^{t+T}\left\Vert \tilde{W}_{c}^{T}\psi\right\Vert ^{2}d\tau\leq-\underline{\psi}\varpi_{7}\left\Vert \tilde{W}_{c}\right\Vert ^{2}+\varpi_{8}\intop_{t}^{t+T}\left\Vert e\right\Vert ^{2}d\tau+3\iota_{2}^{2}\intop_{t}^{t+T}\left\Vert \tilde{W}_{a}\left(\sigma\right)\right\Vert ^{4}d\sigma+\varpi_{9}T,
\]
 where $\varpi_{7}=\frac{\nu^{2}\underline{\varphi}^{2}}{2\left(\nu^{2}\underline{\varphi}^{2}+\eta_{c}^{2}\overline{\varphi}^{2}T^{2}\right)},$
$\varpi_{8}=3\bar{\epsilon^{\prime}}^{2}L_{F}^{2},$ \textup{and $\varpi_{9}=2\left(\iota_{5}^{2}+\bar{\epsilon^{\prime}}^{2}L_{F}^{2}d^{2}\right).$}\end{lem}
\begin{IEEEproof}
The integrand on the LHS can be written as
\[
\tilde{W}_{c}^{T}\left(\tau\right)\psi\left(\tau\right)=\tilde{W}_{c}^{T}\left(t\right)\psi\left(\tau\right)+\left(\tilde{W}_{c}^{T}\left(\tau\right)-\tilde{W}_{c}^{T}\left(t\right)\right)\psi\left(\tau\right).
\]
 Using the inequality $\left(x+y\right)^{2}\geq\frac{1}{2}x^{2}-y^{2}$
and integrating,
\[
\intop_{t}^{t+T}\left(\tilde{W}_{c}^{T}\left(\tau\right)\psi\left(\tau\right)\right)^{2}d\tau\geq\frac{1}{2}\tilde{W}_{c}^{T}\left(t\right)\left(\intop_{t}^{t+T}\left(\psi\left(\tau\right)\psi\left(\tau\right)^{T}\right)d\tau\right)\tilde{W}_{c}\left(t\right)-\intop_{t}^{t+T}\left(\left(\intop_{t}^{\tau}\dot{\tilde{W}}_{c}\left(\sigma\right)d\tau\right)^{T}\psi\left(\tau\right)\right)^{2}d\tau.
\]
 Substituting the dynamics for $\tilde{W}_{c}$ from (\ref{eq:WcDyn})
and using the PE condition in Assumption \ref{APE},
\begin{multline*}
\intop_{t}^{t+T}\left(\tilde{W}_{c}^{T}\left(\tau\right)\psi\left(\tau\right)\right)^{2}d\tau\geq\frac{1}{2}\underline{\psi}\tilde{W}_{c}^{T}\left(t\right)\tilde{W}_{c}\left(t\right)-\intop_{t}^{t+T}\Biggl(\biggl(\intop_{t}^{\tau}\biggl(-\eta_{c}\Gamma\left(\sigma\right)\psi\left(\sigma\right)\psi^{T}\left(\sigma\right)\tilde{W}_{c}\left(\sigma\right)\\
+\frac{\eta_{c}\Gamma\left(\sigma\right)\psi\left(\sigma\right)\Delta\left(\sigma\right)}{\sqrt{1+\nu\omega\left(\sigma\right)^{T}\Gamma\left(\sigma\right)\omega\left(\sigma\right)}}+\frac{\eta_{c}\Gamma\left(\sigma\right)\psi\left(\sigma\right)\tilde{W}_{a}^{T}\mathcal{G}_{\sigma}\tilde{W}_{a}}{4\sqrt{1+\nu\omega\left(\sigma\right)^{T}\Gamma\left(\sigma\right)\omega\left(\sigma\right)}}-\frac{\eta_{c}\Gamma\left(\sigma\right)\psi\left(\sigma\right)\epsilon^{'}\left(\sigma\right)F\left(\sigma\right)}{\sqrt{1+\nu\omega\left(\sigma\right)^{T}\Gamma\left(\sigma\right)\omega\left(\sigma\right)}}\biggl)d\sigma\biggl)^{T}\psi\left(\tau\right)\Biggl)^{2},
\end{multline*}
where $\Delta\triangleq\frac{1}{4}\epsilon'\mathcal{G}\epsilon'^{T}+\frac{1}{2}W^{T}\sigma'\mathcal{G}\epsilon'^{T}.$
Using the inequality $\left(x+y+w-z\right)^{2}\leq2x^{2}+6y^{2}+6w^{2}+6z^{2}$,
\begin{multline*}
\intop_{t}^{t+T}\left(\tilde{W}_{c}^{T}\left(\tau\right)\psi\left(\tau\right)\right)^{2}d\tau\geq\frac{1}{2}\underline{\psi}\tilde{W}_{c}^{T}\left(t\right)\tilde{W}_{c}\left(t\right)-\intop_{t}^{t+T}2\left(\intop_{t}^{\tau}\eta_{c}\tilde{W}_{c}^{T}\left(\sigma\right)\psi\left(\sigma\right)\psi^{T}\left(\sigma\right)\Gamma^{T}\left(\sigma\right)\psi\left(\tau\right)d\sigma\right)^{2}d\tau\\
-6\intop_{t}^{t+T}\left(\intop_{t}^{\tau}\frac{\eta_{c}\Delta^{T}\left(\sigma\right)\psi^{T}\left(\sigma\right)\Gamma^{T}\left(\sigma\right)\psi\left(\tau\right)}{\sqrt{1+\nu\omega\left(\sigma\right)^{T}\Gamma\left(\sigma\right)\omega\left(\sigma\right)}}d\sigma\right)^{2}d\tau-6\intop_{t}^{t+T}\left(\intop_{t}^{\tau}\frac{\eta_{c}F^{T}\left(\sigma\right)\epsilon^{'T}\left(\sigma\right)\psi^{T}\left(\sigma\right)\Gamma^{T}\left(\sigma\right)\psi\left(\tau\right)}{\sqrt{1+\nu\omega\left(\sigma\right)^{T}\Gamma\left(\sigma\right)\omega\left(\sigma\right)}}d\sigma\right)^{2}d\tau\\
-6\intop_{t}^{t+T}\left(\intop_{t}^{\tau}\frac{\eta_{c}\tilde{W}_{a}^{T}\left(\sigma\right)\mathcal{G}_{\sigma}\left(\sigma\right)\tilde{W}_{a}\left(\sigma\right)\psi^{T}\left(\sigma\right)\Gamma^{T}\left(\sigma\right)\psi\left(\tau\right)}{\sqrt{1+\nu\omega\left(\sigma\right)^{T}\Gamma\left(\sigma\right)\omega\left(\sigma\right)}}d\sigma\right)^{2}d\tau.
\end{multline*}
Using the Cauchy-Schwarz inequality, the Lipschitz property, the fact
that $\frac{1}{\sqrt{1+\nu\omega^{T}\Gamma\omega}}\leq1$, and the
bounds in (\ref{eq:bounds}),
\begin{multline*}
\intop_{t}^{t+T}\left(\tilde{W}_{c}^{T}\left(\tau\right)\psi\left(\tau\right)\right)^{2}d\tau\geq\frac{1}{2}\underline{\psi}\tilde{W}_{c}^{T}\left(t\right)\tilde{W}_{c}\left(t\right)-\intop_{t}^{t+T}2\eta_{c}^{2}\left(\intop_{t}^{\tau}\left(\tilde{W}_{c}^{T}\left(\sigma\right)\psi\left(\sigma\right)\right)^{2}d\sigma\intop_{t}^{\tau}\left(\psi^{T}\left(\sigma\right)\Gamma^{T}\left(\sigma\right)\psi\left(\tau\right)\right)^{2}d\sigma\right)d\tau\\
-6\intop_{t}^{t+T}\left(\intop_{t}^{\tau}\frac{\eta_{c}\iota_{5}\overline{\varphi}}{\nu\underline{\varphi}}d\sigma\right)^{2}d\tau-\intop_{t}^{t+T}6\eta_{c}^{2}\iota_{2}^{2}\left(\intop_{t}^{\tau}\left\Vert \tilde{W}_{a}\left(\sigma\right)\right\Vert ^{4}d\sigma\intop_{t}^{\tau}\left(\psi^{T}\left(\sigma\right)\Gamma^{T}\left(\sigma\right)\psi\left(\tau\right)\right)^{2}d\sigma\right)d\tau\\
-\intop_{t}^{t+T}6\eta_{c}^{2}\bar{\epsilon'}^{2}\left(\intop_{t}^{\tau}\left\Vert F\left(\sigma\right)\right\Vert ^{2}d\sigma\intop_{t}^{\tau}\left(\psi^{T}\left(\sigma\right)\Gamma^{T}\left(\sigma\right)\psi\left(\tau\right)\right)^{2}d\sigma\right)d\tau.
\end{multline*}
Rearranging,
\begin{multline*}
\intop_{t}^{t+T}\left(\tilde{W}_{c}^{T}\left(\tau\right)\psi\left(\tau\right)\right)^{2}d\tau\geq\frac{1}{2}\underline{\psi}\tilde{W}_{c}^{T}\left(t\right)\tilde{W}_{c}\left(t\right)-2\eta_{c}^{2}A^{4}\overline{\varphi}^{2}\intop_{t}^{t+T}\left(\tau-t\right)\intop_{t}^{\tau}\left(\tilde{W}_{c}^{T}\left(\sigma\right)\psi\left(\sigma\right)\right)^{2}d\sigma d\tau-3\eta_{c}^{2}A^{4}\overline{\varphi}^{2}\iota_{5}^{2}T^{3}\\
-6\eta_{c}^{2}\iota_{2}^{2}A^{4}\overline{\varphi}^{2}\intop_{t}^{t+T}\left(\tau-t\right)\intop_{t}^{\tau}\left\Vert \tilde{W}_{a}\left(\sigma\right)\right\Vert ^{4}d\sigma d\tau-6\eta_{c}^{2}\bar{\epsilon'}^{2}L_{F}^{2}A^{4}\overline{\varphi}^{2}\intop_{t}^{t+T}\left(\tau-t\right)\intop_{t}^{\tau}\left\Vert e\right\Vert ^{2}d\sigma d\tau-3\eta_{c}^{2}A^{4}\overline{\varphi}^{2}\bar{\epsilon'}^{2}L_{F}^{2}d^{2}T^{3},
\end{multline*}
where $A=\frac{1}{\sqrt{\nu\underline{\varphi}}}$. Changing the order
of integration,
\begin{multline*}
\intop_{t}^{t+T}\left(\tilde{W}_{c}^{T}\left(\tau\right)\psi\left(\tau\right)\right)^{2}d\tau\geq\frac{1}{2}\underline{\psi}\tilde{W}_{c}^{T}\left(t\right)\tilde{W}_{c}\left(t\right)-\eta_{c}^{2}A^{4}\overline{\varphi}^{2}T^{2}\intop_{t}^{t+T}\left(\tilde{W}_{c}^{T}\left(\sigma\right)\psi\left(\sigma\right)\right)^{2}d\sigma\\
-3\eta_{c}^{2}A^{4}\overline{\varphi}^{2}\bar{\epsilon'}^{2}L_{F}^{2}T^{2}\intop_{t}^{t+T}\left\Vert e\left(\sigma\right)\right\Vert ^{2}d\sigma-3\eta_{c}^{2}\iota_{2}^{2}A^{4}\overline{\varphi}^{2}T^{2}\intop_{t}^{t+T}\left\Vert \tilde{W}_{a}\left(\sigma\right)\right\Vert ^{4}d\sigma-2\eta_{c}^{2}A^{4}\overline{\varphi}^{2}T^{3}\left(\iota_{5}^{2}+\bar{\epsilon'}^{2}L_{F}^{2}d^{2}\right).
\end{multline*}
Reordering the terms, the inequality in Lemma 4 is obtained.%

\end{IEEEproof}

\subsection{Gain conditions and gain selection\label{sub:Gain-conditions-and}}

The following section details the procedure to select the control
gains. To facilitate the discussion, define $\eta_{a12}\triangleq\eta_{a1}+\eta_{a2}$,
$Z\triangleq\begin{bmatrix}e^{T} & \tilde{W}_{c}^{T} & \tilde{W}_{a}^{T}\end{bmatrix}^{T},$
$\iota\triangleq\frac{\left(\eta_{a2}\overline{W}+\iota_{4}\right)^{2}}{\eta_{a12}}+2\eta_{c}\left(\iota_{1}\right)^{2}+\frac{1}{4}\iota_{3},$
$\varpi_{10}\triangleq\frac{\varpi_{6}\eta_{a12}+2\varpi_{2}\underline{q}+\eta_{c}\varpi_{9}}{8}+\iota$,
$\varpi_{11}\triangleq\frac{1}{16}\min(\eta_{c}\underline{\psi}\varpi_{7},\:2\varpi_{0}\underline{q}T,\:\varpi_{3}\eta_{a12}T),$
$Z_{0}\in\mathbb{R}_{\geq0}$ denotes a known constant bound on the
initial condition such that $\left\Vert Z\left(t_{0}\right)\right\Vert \leq Z_{0}$,
and 
\begin{align}
\overline{Z} & \triangleq\underline{v_{l}}^{-1}\left(\overline{v_{l}}\left(\max\left(\left\Vert Z_{0}\right\Vert ^{2},\frac{\varpi_{10}T}{\varpi_{11}}\right)\right)+\iota T\right).\label{eq:Zbar}
\end{align}
The sufficient gain conditions for the subsequent Theorem \ref{thm:main_thm}
are given by 
\begin{gather}
\eta_{a12}>\max\left(\eta_{a1}\xi_{2}+\frac{\eta_{c}\iota_{2}}{4}\sqrt{\frac{\overline{Z}}{\nu\underline{\varphi}}},3\eta_{c}\iota_{2}^{2}\overline{Z}\right),\nonumber \\
\xi_{1}>2\overline{\epsilon^{\prime}}L_{F},\:\eta_{c}>\frac{\eta_{a1}}{\lambda\underline{\gamma}\xi_{2}},\:\underline{\psi}>\frac{2\varpi_{4}\eta_{a12}}{\eta_{c}\varpi_{7}}T,\nonumber \\
\underline{q}>\max\left(\frac{\varpi_{5}\eta_{a12}}{\varpi_{0}},\frac{1}{2}\eta_{c}\varpi_{8},\eta_{c}L_{F}\overline{\epsilon^{\prime}}\xi_{1}\right),\nonumber \\
T<\min\Biggl(\frac{1}{\sqrt{6N}\eta_{a12}},\frac{\nu\underline{\varphi}}{\sqrt{6N}\eta_{c}\overline{\varphi}},\frac{1}{2\sqrt{n}L_{F}},\sqrt{\frac{\eta_{a12}}{6N\eta_{a12}^{3}+8\underline{q}\varpi_{1}}}\Biggr).\label{eq:SuffCond}
\end{gather}
Since the gains depend on the initial conditions and on the compact
sets used for function approximation and Lipschitz bounds, an iterative
algorithm is developed to select the gains.
\begin{algorithm}
\uline{First iteration:}

Given $Z_{0}\in\mathbb{R}_{\geq0}$ such that $\left\Vert Z\left(t_{0}\right)\right\Vert <Z_{0}$,
let $\mathcal{Z}_{1}=\left\{ \varrho\in\mathbb{R}^{n+2\left\{ N\right\} _{1}}\mid\left\Vert \varrho\right\Vert \leq\beta_{1}\underline{v_{l}}^{-1}\left(\overline{v_{l}}\left(\left\Vert Z_{0}\right\Vert \right)\right)\right\} $
for some $\beta_{1}>1$. Using $\mathcal{Z}_{1},$ compute the bounds
in (\ref{eq:bounds}) and (\ref{eq:Zbar}), and select the gains according
to (\ref{eq:SuffCond}). If $\left\{ \overline{Z}\right\} _{1}\leq\beta_{1}\underline{v_{l}}^{-1}\left(\overline{v_{l}}\left(\left\Vert Z_{0}\right\Vert \right)\right),$
set $\mathcal{Z}=\mathcal{Z}_{1}$ and terminate.

\uline{Second iteration:}

If $\left\{ \overline{Z}\right\} _{1}>\beta_{1}\underline{v_{l}}^{-1}\left(\overline{v_{l}}\left(\left\Vert Z_{0}\right\Vert \right)\right),$
let $\mathcal{Z}_{2}\triangleq\left\{ \varrho\in\mathbb{R}^{n+2\left\{ N\right\} _{1}}\mid\left\Vert \varrho\right\Vert \leq\beta_{2}\left\{ \overline{Z}\right\} _{1}\right\} $.
Using $\mathcal{Z}_{2},$ compute the bounds in (\ref{eq:bounds})
and (\ref{eq:Zbar}) and select the gains according to (\ref{eq:SuffCond}).
If $\left\{ \overline{Z}\right\} _{2}\leq\left\{ \overline{Z}\right\} _{1}$,
set $\mathcal{Z}=\mathcal{Z}_{2}$ and terminate.

\uline{Third iteration:}

If $\left\{ \overline{Z}\right\} _{2}>\left\{ \overline{Z}\right\} _{1}$,
increase the number of NN neurons to $\left\{ N\right\} _{3}$ to
yield a lower function approximation error $\left\{ \overline{\epsilon^{\prime}}\right\} _{3}$
such that $\left\{ L{}_{F}\right\} _{2}\left\{ \overline{\epsilon^{\prime}}\right\} _{3}\leq\left\{ L{}_{F}\right\} _{1}\left\{ \overline{\epsilon^{\prime}}\right\} _{1}$.
The increase in the number of NN neurons ensures that $\left\{ \iota\right\} _{3}\leq\left\{ \iota\right\} _{1}$.
Furthermore, the assumption that the PE interval $\left\{ T\right\} _{3}$
is small enough such that $\left\{ L_{F}\right\} _{2}\left\{ T\right\} _{3}\leq\left\{ T\right\} _{1}\left\{ L_{F}\right\} _{1}$
and $\left\{ N\right\} _{3}\left\{ T\right\} _{3}\leq\left\{ T\right\} _{1}\left\{ N\right\} _{1}$
ensures that $\left\{ \frac{\varpi_{10}}{\varpi_{11}}\right\} _{3}\leq\left\{ \frac{\varpi_{10}}{\varpi_{11}}\right\} _{1}$,
and hence, $\left\{ \overline{Z}\right\} _{3}\leq\beta_{2}\left\{ \mbox{\ensuremath{\overline{Z}}}\right\} _{1}$.
Set $\mathcal{Z}=\left\{ \varrho\in\mathbb{R}^{n+2\left\{ N\right\} _{3}}\mid\left\Vert \varrho\right\Vert \leq\beta_{2}\left\{ \overline{Z}\right\} _{1}\right\} $
and terminate.

\caption{\label{alg:Gain-Selection}Gain Selection}
\end{algorithm}
 In Algorithm \ref{alg:Gain-Selection}, the notation $\left\{ \varpi\right\} _{i}$
for any parameter $\varpi$ denotes the value of $\varpi$ computed
in the $i^{th}$ iteration. Algorithm \ref{alg:Gain-Selection} ensures
that the selected compact set $\mathcal{Z}$ satisfies $\overline{Z}\in\mathcal{Z}$.
\begin{thm}
\label{thm:main_thm}Provided that the sufficient conditions in (\ref{eq:SuffCond})
are satisfied and Assumptions \ref{Apsinv} - \ref{APE} hold, the
controller in (\ref{eq:control}) and the update laws in (\ref{eq:WcHatdot})
- (\ref{eq:WaHatdot}) guarantee that the tracking error is ultimately
bounded, and the error between the policy $\mu$ and the optimal policy
$\mu^{*}$ is ultimately bounded.\end{thm}
\begin{IEEEproof}
Consider the candidate Lyapunov function $V_{L}:\mathbb{R}^{n+2N}\times\mathbb{R}_{\geq0}\to\mathbb{R}$
defined as $V_{L}\left(Z,t\right)\triangleq V_{t}^{*}\left(e,t\right)+\frac{1}{2}\tilde{W}_{c}^{T}\Gamma^{-1}\tilde{W}_{c}+\frac{1}{2}\tilde{W}_{a}^{T}\tilde{W}_{a}.$
Using Lemma \ref{lem:V*pd} and (\ref{eq:GammaBound}), 
\begin{equation}
\underline{v_{l}}\left(\left\Vert Z\right\Vert \right)\leq V_{L}\left(Z,t\right)\leq\overline{v_{l}}\left(\left\Vert Z\right\Vert \right),\label{eq:VLBound}
\end{equation}
$\forall Z\in B_{b},\:\forall t\in\mathbb{R}_{\geq0}$, where $\underline{v_{l}}:\left[0,b\right]\rightarrow\mathbb{R}_{\geq0}$
and $\overline{v_{l}}:\left[0,b\right]\rightarrow\mathbb{R}_{\geq0}$
are class $\mathcal{K}$ functions, and $B_{b}\subset\mathbb{R}^{n+2N}$
denotes a ball of radius $b\in\mathbb{R}_{>0}$ around the origin.

The time derivative of $V_{L}$ is $\dot{V}_{L}=V^{*\prime}F+V^{*\prime}G\mu+\tilde{W}_{c}^{T}\Gamma^{-1}\dot{\tilde{W}}_{c}-\frac{1}{2}\tilde{W}_{c}^{T}\Gamma^{-1}\dot{\Gamma}\Gamma^{-1}\tilde{W}_{c}-\tilde{W}_{a}^{T}\dot{\hat{W}}_{a}.$
Using (\ref{eq:WcDyn}) and the facts that $V^{*\prime}F=-V^{*\prime}G\mu^{*}-r\left(\zeta,\mu^{*}\right)$
and $V^{*\prime}G=-2\mu^{*T}R$ yields
\begin{multline}
\dot{V}_{L}=-e^{T}Qe+\mu^{*T}R\mu^{*}-2\mu^{*T}R\mu-\eta_{c}\tilde{W}_{c}^{T}\psi\psi^{T}\tilde{W}_{c}-\lambda\frac{\eta_{C}}{2}\tilde{W}_{c}^{T}\Gamma^{-1}\tilde{W}_{c}+\frac{1}{2}\eta_{c}\tilde{W}_{c}^{T}\frac{\omega\omega^{T}}{\rho}\tilde{W}_{c}-\tilde{W}_{a}^{T}\dot{\hat{W}}_{a}\\
+\frac{\eta_{c}\tilde{W}_{c}^{T}\psi}{\sqrt{1+\nu\omega^{T}\Gamma\omega}}\biggl(\frac{1}{4}\tilde{W}_{a}^{T}\mathcal{G}_{\sigma}\tilde{W}_{a}-\epsilon^{\prime}F+\frac{1}{4}\epsilon^{\prime}\mathcal{G}\epsilon^{\prime T}+\frac{1}{2}W^{T}\sigma^{\prime}\mathcal{G}\epsilon^{\prime T}\biggr),\label{eq:VlDot2}
\end{multline}
where $\rho\triangleq1+\nu\omega^{T}\Gamma\omega$. Using (\ref{eq:WcHatdot}),
(\ref{eq:deltaUnm}) and the bounds in (\ref{eq:psibound}) - (\ref{eq:bounds})
the Lyapunov derivative in (\ref{eq:VlDot2}) can be bounded above
on the set $\mathcal{Z}$ as 
\begin{multline*}
\dot{V}_{L}\leq-\frac{\underline{q}}{2}\left\Vert e\right\Vert ^{2}-\frac{1}{4}\eta_{c}\left\Vert \tilde{W}_{c}^{T}\psi\right\Vert ^{2}-\frac{\eta_{a12}}{2}\left\Vert \tilde{W}_{a}\right\Vert ^{2}+\left(2\eta_{a2}\overline{W}+\iota_{4}\right)\left\Vert \tilde{W}_{a}\right\Vert +\eta_{c}\left(\iota_{1}+\iota_{2}\overline{W}^{2}\right)\left\Vert \tilde{W}_{c}^{T}\psi\right\Vert \\
-\frac{1}{2}\left(\eta_{a12}-\eta_{a1}\xi_{2}-\frac{\eta_{c}\iota_{2}}{4}\left\Vert \tilde{W}_{c}^{T}\psi\right\Vert \right)\left\Vert \tilde{W}_{a}\right\Vert ^{2}-\frac{1}{2}\eta_{c}\left(1-\frac{\overline{\epsilon^{\prime}}}{\xi_{1}}\right)\left\Vert \tilde{W}_{c}^{T}\psi\right\Vert ^{2}-\frac{1}{2}\left(\underline{q}-\eta_{c}L_{F}\overline{\epsilon^{\prime}}\xi_{1}\right)\left\Vert e\right\Vert ^{2}\\
-\frac{1}{2}\left(\lambda\eta_{c}\underline{\gamma}-\frac{\eta_{a1}}{\xi_{2}}\right)\left\Vert \tilde{W}_{c}\right\Vert ^{2}+\frac{1}{4}\iota_{3},
\end{multline*}
where $\xi_{1}$, $\xi_{2}\in\mathbb{R}$ are known adjustable positive
constants. Provided the sufficient conditions in (\ref{eq:SuffCond})
are satisfied, completion of squares yields %
\begin{equation}
\dot{V}_{L}\leq-\frac{\underline{q}}{2}\left\Vert e\right\Vert ^{2}-\frac{1}{8}\eta_{c}\left\Vert \tilde{W}_{c}^{T}\psi\right\Vert ^{2}-\frac{\eta_{a12}}{4}\left\Vert \tilde{W}_{a}\right\Vert ^{2}+\iota.\label{eq:VlDot4}
\end{equation}
The inequality in (\ref{eq:VlDot4}) is valid provided $Z\left(t\right)\in\mathcal{Z}$.
Integrating (\ref{eq:VlDot4}) and using Lemma \ref{lem:WcBound}
and the gain conditions in (\ref{eq:SuffCond}) yields 
\[
V_{L}\left(Z\left(t+T\right),t+T\right)-V_{L}\left(Z\left(t\right),t\right)\leq-\frac{1}{8}\eta_{c}\underline{\psi}\varpi_{7}\left\Vert \tilde{W}_{c}\left(t\right)\right\Vert ^{2}-\frac{\underline{q}}{4}\intop_{t}^{t+T}\left\Vert e\left(\tau\right)\right\Vert ^{2}d\tau-\frac{\eta_{a12}}{8}\intop_{t}^{t+T}\left\Vert \tilde{W}_{a}\left(\tau\right)\right\Vert ^{2}d\tau+\frac{1}{8}\eta_{c}\varpi_{9}+\iota T,
\]
provided $Z\left(\tau\right)\in\mathcal{Z},\:\forall\tau\in\left[t,t+T\right]$.
Using the facts that $-\intop_{t}^{t+T}\left\Vert e\left(\tau\right)\right\Vert ^{2}d\tau\leq-T\inf_{\tau\in\left[t,t+T\right]}\left\Vert e\left(\tau\right)\right\Vert ^{2}$
and $-\intop_{t}^{t+T}\left\Vert \tilde{W}_{a}\left(\tau\right)\right\Vert ^{2}d\tau\leq-T\inf_{\tau\in\left[t,t+T\right]}\left\Vert \tilde{W}_{a}\left(\tau\right)\right\Vert ^{2}$,
and Lemma \ref{lem:WaBound} yield%
\[
V_{L}\left(Z\left(t+T\right),t+T\right)-V_{L}\left(Z\left(t\right),t\right)\leq-\frac{\eta_{c}\underline{\psi}\varpi_{7}}{16}\left\Vert \tilde{W}_{c}\left(t\right)\right\Vert ^{2}-\frac{\varpi_{0}\underline{q}T}{8}\left\Vert e\left(t\right)\right\Vert ^{2}-\frac{\varpi_{3}\eta_{a12}T}{16}\left\Vert \tilde{W}_{a}\left(t\right)\right\Vert ^{2}+\varpi_{10}T,
\]
provided $Z\left(\tau\right)\in\mathcal{Z},\:\forall\tau\in\left[t,t+T\right]$.
Thus, $V_{L}\left(Z\left(t+T\right),t+T\right)-V_{L}\left(Z\left(t\right),t\right)<0$
provided $\left\Vert Z\left(t\right)\right\Vert >\frac{\varpi_{10}T}{\varpi_{11}}$
and $Z\left(\tau\right)\in\mathcal{Z},\forall\tau\in\left[t,t+T\right]$.
The bounds on the Lyapunov function in (\ref{eq:VLBound}) yield $V_{L}\left(Z\left(t+T\right),t+T\right)-V_{L}\left(Z\left(t\right),t\right)<0$
provided $V_{L}\left(Z\left(t\right),t\right)>\overline{v_{l}}\left(\frac{\varpi_{10}T}{\varpi_{11}}\right)$
and $Z\left(\tau\right)\in\mathcal{Z},\:\forall\tau\in\left[t,t+T\right]$. 

Since $Z\left(t_{0}\right)\in\mathcal{Z},$ (\ref{eq:VlDot4}) can
be used to conclude that $\dot{V}_{L}\left(Z\left(t_{0}\right),t_{0}\right)\leq\iota$.
The iterative gain selection procedure in Algorithm \ref{alg:Gain-Selection}
ensures that $\underline{v_{l}}^{-1}\left(V_{L}\left(Z\left(t_{0}\right),t_{0}\right)+\iota T\right)\leq\overline{Z}$;
and hence, $Z\left(t\right)\in\mathcal{Z}$ for all $t\in[t_{0},t_{0}+T]$.
If $V_{L}\left(Z\left(t_{0}\right),t_{0}\right)>\overline{v_{l}}\left(\frac{\varpi_{10}T}{\varpi_{11}}\right)$,
then $Z\left(t\right)\in\mathcal{Z}$ for all $t\in[t_{0},t_{0}+T]$
implies $V_{L}\left(Z\left(t_{0}+T\right),t_{0}+T\right)-V_{L}\left(Z\left(t_{0}\right),t_{0}\right)<0$.
Thus, the iterative gain selection procedure in Algorithm \ref{alg:Gain-Selection}
ensures that $\underline{v_{l}}^{-1}\left(V_{L}\left(Z\left(t_{0}+T\right),t_{0}+T\right)+\iota T\right)\leq\overline{Z}$;
and hence, $Z\left(t\right)\in\mathcal{Z}$ for all $t\in[t_{0}+T,t_{0}+2T]$.
Inductively, the system state is bounded such that $\sup_{t\in[0,\infty)}\left\Vert Z\left(t\right)\right\Vert ^{2}\leq\overline{Z}$
and ultimately bounded such that 
\[
\lim\sup_{t\to\infty}\left\Vert Z\left(t\right)\right\Vert ^{2}\leq\underline{v_{l}}^{-1}\left(\overline{v_{l}}\left(\frac{\varpi_{10}T}{\varpi_{11}}\right)+\iota T\right).
\]

\end{IEEEproof}

\section{Conclusion}

An ADP-based approach using the policy evaluation and policy improvement
architecture is presented to approximately solve the infinite horizon
optimal tracking problem for control affine nonlinear systems with
quadratic cost. The problem is solved by transforming the system to
convert the tracking problem that has a time-varying value function,
into a time-invariant optimal control problem. The ultimately bounded
tracking and estimation result was established using Lyapunov analysis
for nonautonomous systems. The developed method can be applied to
high-dimensional nonlinear dynamical systems using simple polynomial
basis functions and sinusoidal probing signals. However, the accuracy
of the approximation depends on the choice of basis functions and
the result hinges on the system states being PE. Furthermore, computation
of the desired control in (\ref{eq:ud}) requires exact model knowledge. 

A solution to the tracking problem without using the desired control
and employs a multi-layer neural network that can approximate the
basis functions remains a future challenge. In adaptive control, it
is generally possible to formulate the control problem such that PE
along the desired trajectory is sufficient to achieve parameter convergence.
In the ADP-based tracking problem, PE along the desired trajectory
would be sufficient to achieve parameter convergence if the BE can
be formulated in terms of the desired trajectories. Achieving such
a formulation is not trivial, and is a subject for future research. 

\bibliographystyle{IEEEtran}
\bibliography{ncr,master,encr}

\end{document}